\documentclass[a4paper,twocolumn,superscriptaddress,11pt]{quantumarticle}
\pdfoutput=1

\usepackage[utf8]{inputenc}
\usepackage[english]{babel}
\usepackage[T1]{fontenc}
\usepackage{amsmath}
\usepackage{hyperref}

\usepackage{tikz}
\usepackage{lipsum}

\usepackage[left=0.875in,right=0.875in,top=1.0in,bottom=1.25in]{geometry}
\usepackage{amsfonts,amssymb}
\usepackage{graphicx}

\usepackage[super,numbers]{natbib}
\begin{document}

\title{Well defined quantum key distribution using calibration, synchronization, and a programmable quantum channel}
\date{\today}
\author{Billy E.~Geerhart}
\author{Venkat  R.~Dasari,Ph.D.}
\affil{U.S Army Research Laboratory, 320 Hopkins Road, Aberdeen Proving Ground, MD 21005}
\thanks{Further author information: (Send correspondence to Venkat R.Dasari)\\E-mail: venkateswara.r.dasari.civ@mail.mil\\ Telephone: +01-410-278-2846}

%\email{latex@quantum-journal.org}
%\homepage{http://quantum-journal.org}
%\orcid{0000-0003-0290-4698}
\maketitle

\begin{abstract}
  Well defined quantum key distribution between two users requires both calibration to ensure quantum effects and synchronization to stabilize the bit parity of the results. Here we present two quantum effects regarding two entangled photons in a single fiber that can be used for both calibration and synchronization. In particular, we show how the synchronization problem can be transformed from a maximization of the bit parity between two photons sent to two users to finding an average 50/50 bit parity for two photons sent to a single user; the end result being first order feedback rather than second order feedback. Once we show how to calibrate and synchronize a quantum channel for two users, we then show how to introduce multiple users through a programmable quantum channel that can change its configuration depending on who needs to exchange quantum information. The programmable quantum channel is created by using a programmable classical channel to control the quantum devices as well as introducing new metadata on the classical channel specific to quantum applications.
\end{abstract}

\section{Introduction}\label{sec:intro}

Solutions to multi-user quantum networks in support of quantum applications are being explored today\cite{quantumnetworks2017}. The applications usually involve projects that enhance security\cite{Williams2016}, enhance stability\cite{qkd_timebin1,stability1}, enhance bandwidth\cite{qkd_timebin2}, or enhance the capability of quantum sensors such as full Bell state analysis using linear optics\cite{PhysRevLett.118.050501}. Of particular interest are the challenges facing quantum key distribution(QKD). QKD can be implemented using a weak photon source at Alice to encode information in a random basis which is then sent to Bob\cite{quantumreview2016}. QKD can also be implemented using photons entangled into a Bell state in the middle with one photon sent to each of Alice and Bob\cite{quantumreview2018, dasari2018qkd}. QKD can even be implemented using a photon source at both Alice and Bob which is then projected into an entangled state using Bell state analysis\cite{qkd_mdi,qkd_mdi2}. Each application involves photons traveling between Alice and Bob, so a common problem is that photons traveling through birefringent fiber inherently changes the bit parity of the shared secret. Indeed, one can actually touch a birefringent fiber and the thermal expansion of the fiber will make the bit parity oscillate with a period on the order of seconds. The bit parity depends on the distance traveled between the horizontal photons relative to the vertical photons, as such the bit parity oscillation is caused by thermal expansion of the fiber as this makes the slow axis photons travel further than the fast axis photons. In the worst case scenario the fast and slow axis photons are separated so far that the birefringent fiber itself acts like a measuring device in the zero degree basis; this is the default case when birefringent fibers are connected without proper calibrations. To alleviate these problems, section \ref{sec:Calibration} shows how to calibrate birefringent fibers to maintain quantum effects, while section \ref{sec:Synchronization} shows how to synchronize the quantum channel to maintain a stable bit parity in the face of thermal expansion of the birefringent fibers.

Once the quantum channel can be calibrated and synchronized between two users, a quantum relay can be added to support quantum key distribution between three users. Previously we have proposed using a programmable classical channel to control a quantum channel through the introduction of new metadata on the classical channel as well as standardizing how the quantum devices are configured\cite{dasari2016openflow}. In this case the programmable classical channel can intercept packets sent by one of the hosts, configure the quantum relay to connect two users in preparation for a quantum key exchange, and then the programmable classical channel can inform the end users that the quantum devices are ready for a key exchange. As outlined in section \ref{sec:programmable_control}, the proposed standards are used to merge a programmable classical channel with a configurable quantum channel to create a programmable quantum channel.

\begin{figure}[!htb]
\centering
\includegraphics[width=.95\linewidth]{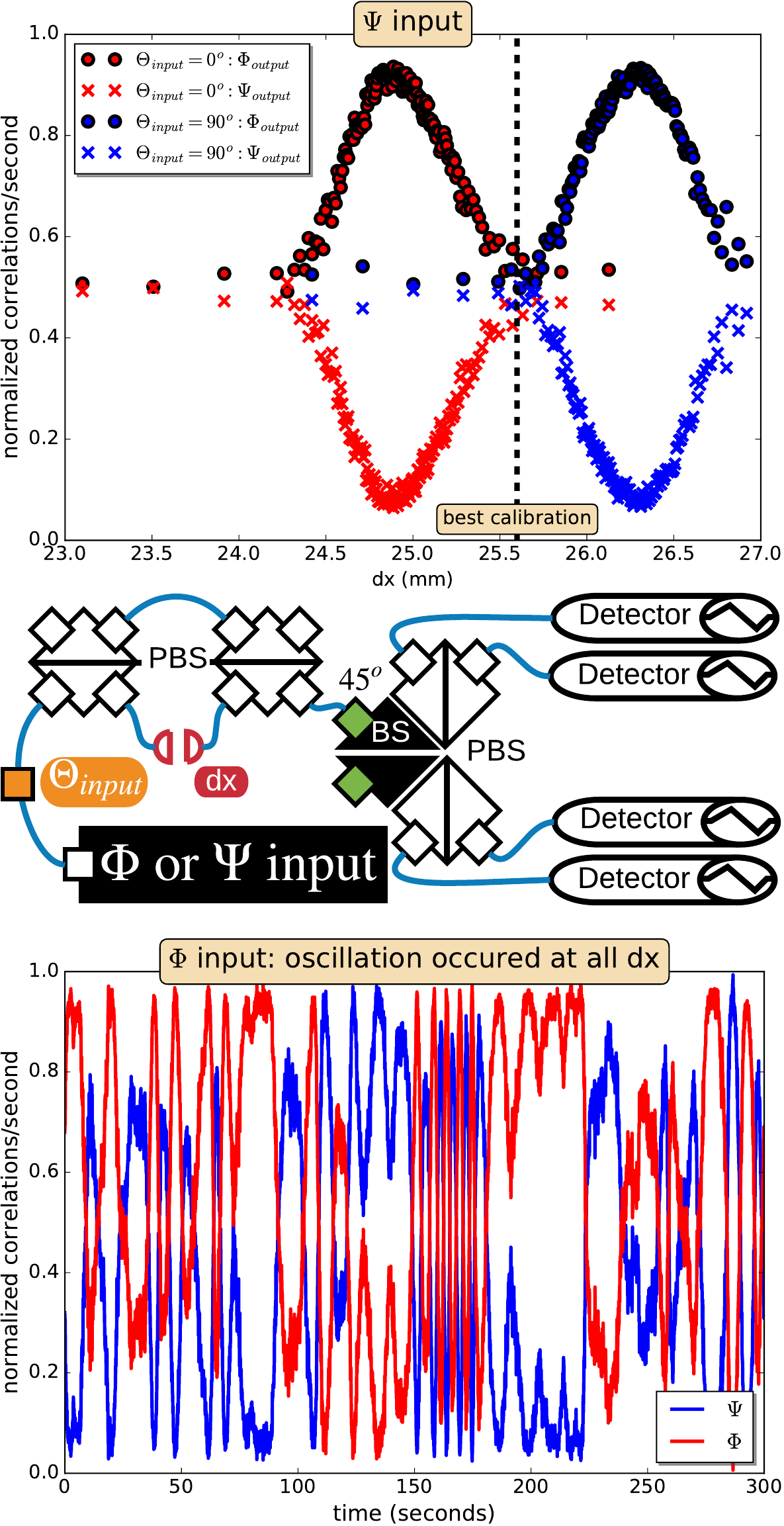}
\caption{The schematic in the center shows how a single fiber can be used as a quantum sensor using two entangled particles inside a single fiber. The top plot shows the signal for an input $\Psi$ state producing a $\Psi$ dip regardless of phase. To find the calibration point, simply scan twice with the input/output being rotated by $90^o$, then the best calibration point is between the two signals. The bottom plot shows the signal for an input $\Phi$ state that oscillates as the path difference between the fast and slow axis varies; this oscillation can be used to sense the thermal expansion of a birefringent fiber.}
\label{fig:single_fiber_sensor}
\end{figure}
\section{Calibration}\label{sec:Calibration}
Many quantum applications requires every birefringent fiber to have a roughly zero path length difference between the fast and slow axis. Calibration can be roughly achieved by splicing the birefringent fiber at the midpoint and reconnecting the ends $90^o$ relative to each other. Another technique is to add two polarizing beam splitters to the fiber and varying the path length difference between the fast and slow axis until quantum effects are observed. A more deterministic approach is to connect a type II photon source to the input side of the birefringent fiber and then connect a $45^o$ measuring device to the output of the birefringent fiber(see fig. \ref{fig:single_fiber_sensor}), next scan across path length differences until quantum effects are observed, and then rotate both the input and output of the optical fiber coupling by $90^o$ and perform another scan; the results of such a procedure can be seen in fig. \ref{fig:single_fiber_sensor} where the best calibration point is actually the mid point between the two quantum signals. This procedure can be performed on each birefringent fiber of a quantum application to ensure quantum effects are observed.
\section{Synchronization}\label{sec:Synchronization}
\begin{figure}[!htb]
\centering
\includegraphics[width=1.\linewidth]{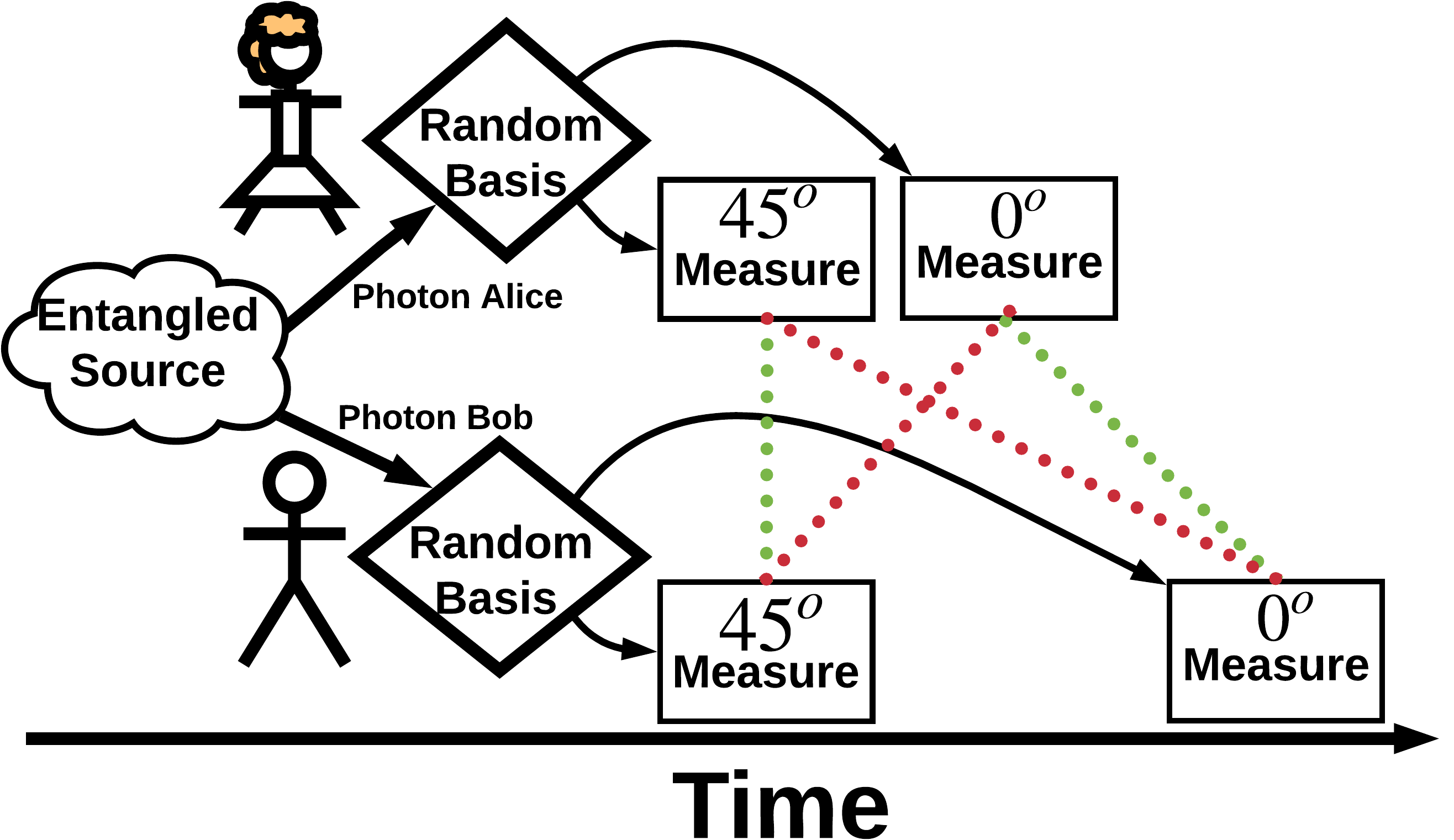}
\caption{This diagram shows how time encoded QKD uses asymmetric delays to encode the measurement basis used between Alice and Bob. Alice and Bob choose to randomly measure in the $45^o$ or $0^o$ basis. An asymmetric delay is added to the $0^o$ basis that allows Alice and Bob to determine which measurement basis was used. The green dotted lines represent a successful key exchange as the basis used by Alice and Bob are the same.}
\label{fig:time_encoded_QKD}
\end{figure}

Synchronization is the process of stabilizing bit parity correlations over time. The details are specific to the quantum application. For instance, in time encoded QKD(see fig. \ref{fig:time_encoded_QKD}) the basis used is encoded in the relative delays of the coincident photons\cite{dasari2018bellstate}. The bit parity itself depends on the phase of the received photons; this phase is directly proportional to how far the photon pair has traveled before being randomly handed out to the end users, but after being handed out the phase is proportional to the relative distance between the fast and slow axis for a single birefringent fiber. The input side is stabilized by adding a beam splitters as a filter\cite{beamsplitters1, beamsplitters2} that allows the source to consistently share $\Psi^-$ states between two users, and the unwanted $\Psi^+$ states are converted to $\Phi^+$ states which are filtered to a single random user. Even though the source is stabilized, the output phase will oscillate when the slow and fast axis path length differences change. Unfortunately this occurs just by thermal expansion of the birefringent fiber. To compensate, the end users can perform quantum sensing on the filtered $\Phi^-$ state. Figure \ref{fig:single_fiber_sensor} shows the bit parity oscillates as the filtered $\Phi$ state oscillates between $\Phi^+$ and $\Phi^-$. Knowing we have an input filtered $\Phi^-$ state, we can fix the path length difference of a single birefringent fiber by adding a liquid crystal to prevent filtered $\Phi$ states from oscillating in bit parity. The main difficulty of this method can be seen if we consider the case of one users fiber staying constant while the other users fiber undergoes thermal expansion, in this scenario the filtered $\Phi^-$ state changes phase twice as fast as the shared $\Psi^-$ state because both photons of the filtered state exist in the single fiber undergoing thermal expansion. Although being a maximization problem on the filtered $\Phi^-$ states, an actual implementation would have to maximize the filtered $\Phi^-$ bit parity and then checking the shared $\Psi^-$ bit parity results; the worst case scenario being one user has to cycle by half a wave length on the filtered $\Phi^-$ states to maximize the bit parity on the shared $\Psi^-$ states. Adding a filter on the source allowed us to stabilize the source, while using quantum sensing on a single fiber allows us to stabilize the birefringent fibers going to the end users.

Synchronization has been reduced to using a passive filter on the source and using maximization of the bit parity on the filtered $\Phi^-$ states , but we can actually transform the maximization problem to a split 50/50 bit parity problem through the introduction of $\pi/4$ wave plates on the slow axis of each end user. The wave-plates convert the filtered $\Phi^-$ states to an equal superposition of filtered $\Phi^+$ and $\Phi^-$ states which corresponds with a 50/50 bit parity, yet the bit parity between the shared $\Psi^-$ state remains unchanged as both terms in the $\Psi^-$ state experience the same phase shift from the wave plates. The procedure is exactly the same as before: each user searches for an applied phase that produces a 50/50 bit parity on the filtered $\Phi^-$ states while double checking the bit parity on the shared $\Psi^-$ states. Since we are already searching for an applied phase that produces a 50/50 bit parity on the filtered $\Phi^-$ states, we can simply omit the 1/4 wave-plates and instead use variable liquid crystals so long as each end user aligns the wave plate in the same fast or slow axis. By transforming the problem from a maximization problem to a 50/50 bit parity problem, we avoid the feedback problems regarding the maximization problem. In particular the photon correlation counts will give a first order response rather than a second order response.
\section{Programmable Control}\label{sec:programmable_control}
Programmable control has already been demonstrated on the classical channel using such standards as Openflow\cite{de2014using}. A typical switch will have the data plane unified with the control plane on a single device. The data plane itself is the fast switching operations, while the control plane is any protocols that are put in place to deal with new connections or hosts that appear on the network. Openflow separates the control plane from the data plane, and then puts in place abstractions necessary for each functional unit to be developed separately. The data plane is the fast switching associated with a typical switch, except in the Openflow standard the switching is modified to include flow tables that have match conditions and instructions. The new flow tables allows for simple logic to be applied on the fast switching data plane. These flow tables usually exist on a physical switch that implements the Openflow standard as it relates to the data plane, while the control plane is separated on another computing device called a controller. The controller can communicate with the data plane to update its flow tables, or the data plane can communicate with the controller whenever it receives a packet that matches on an instruction that specifically says to forward to the controller. Although the controller is slower than the data plane in terms of routing, the controller is where more complex logic is applied to determine what to do with packets that currently don't fit within the data planes forwarding rules. Once a decision is made, the controller can then add additional flow table entries to the data plane based on what it learns from the packet it received. The essential point is the classical channel becomes programmable with the introduction of an Openflow switch and an Openflow controller that can be programmed to populate the flow tables of the switch.

To create a programmable quantum channel, we can merge the programmable classical channel with a configurable quantum channel\cite{dasari2016programmable}. Previously we have proposed abstractions which allow a programmable classical channel to control a configurable quantum channel\cite{dasari2017abstractions}. To demonstrate these abstractions, consider time encoded QKD wherein asymmetric delays are added to encode the basis used between two users (see fig \ref{fig:time_encoded_QKD}). The asymmetric delays can actually be hard coded for two users, but a third user would need a configurable delay that could change depending on who this third person was connected with. One of the proposed abstractions consists of treating the quantum devices as configurable net devices. The quantum device is then equipped with a NETCONF server that communicates with a NETCONF client using the YANG data model. The client can communicate with the server to retrieve information from the quantum device or to send configuration updates. Once the quantum device is equipped with a NETCONF server, the programmable classical channel can then configure the delays in a way that is vendor agnostic. After standardizing the interface to the quantum device, the next abstraction necessary for programmable control is metadata on the classical channel associated with quantum applications. The proposed metadata includes QCHANNEL for quantum channel, QCOM for quantum communication protocol, and QEC for quantum error correction protocol. This metadata is included on packets on the programmable classical channel to allow the programmable classical channel to react to communications specific to quantum applications. For instance, Charlie can request a quantum key exchange to Alice using the new quantum metadata. The programmable switch will intercept the packet based on the new metadata and forward the request to the controller. Next the controller can respond by requesting time slots from the relay to connect Alice and Charlie together, and then the controller can send this information to both Alice and Charlie while also reconfiguring Charlie's asymmetric delay to be compatible with Alice's delay. By treating each quantum device as a configurable net device, the programmable classical channel can control the quantum channel; by adding additional metadata specific to quantum applications, the programmable classical channel can be integrated with the quantum channel to create a programmable quantum channel.
\section{Results and Discussion} \label{sec:results}

\begin{figure}[htb]
\centering
\includegraphics[width=1.\linewidth]{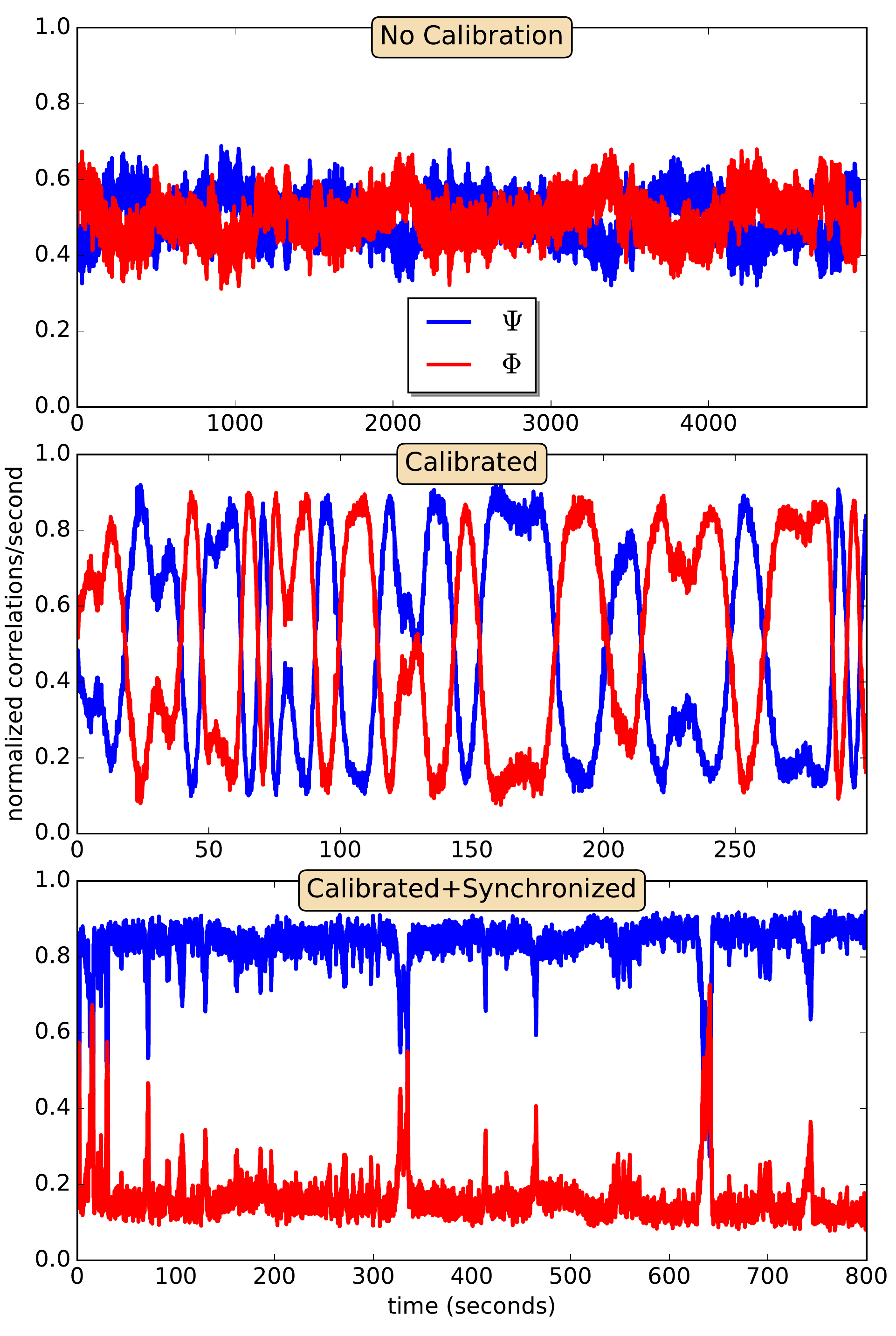}
%\vspace{-9.8cm}
\caption{These three graphs show the benefits of calibration and synchronization. All three graphs show the normalized correlations/second for an input $\Psi$ entangled state with a $45^o$ measurement on the output. With no calibration there are no quantum effects. With calibration there are quantum effects, but the phase shifts over time. With calibration and synchronization the bit parity is controlled.}
\label{fig:finalData}
\end{figure}
The calibration procedure was tested using a $\Psi$ state measured in the $45^o$ basis. Before calibration the bit parity is purely random(see fig. \ref{fig:finalData}). After the calibration procedure outlined in section \ref{sec:Calibration} was performed, the quantum effects produce a time varying bit parity. The time variation is caused by the state oscillating between $\Psi^-$ and $\Psi^+$. The synchronization procedure performed for the calibrated+synchronized plot is just a maximization of the bit parity using active feedback. The synchronization was difficult because the observed bit parity is a secondary response to the applied phase, which is to say that near maximization a large change in phase will produce a small change in bit parity. This maximization problem is what inspired the synchronization procedure outlined in section \ref{sec:Synchronization} in which the maximization problem is transformed into a 50/50 bit parity problem for two photons sent to a single random user.
\section{Conclusion} \label{sec:conclusion}
We presented two quantum effects which can be used for calibration and synchronization. Both quantum effects relied on sending two entangled photons down a single fiber followed by a measurement in the $45^o$ basis. Varying the calibration, an input $\Psi$ state produced a consistent dip regardless of phase, while an input $\Phi$ state produced a bit parity that oscillated depending on how far apart the H and V photons were separated. It's interesting to note the input $\Phi$ oscillation was equally strong over all path length differences tested(range of 3cm). The $\Psi$ dip can be used for the calibration procedure outlined in section \ref{sec:Calibration}, while the $\Phi$ oscillation can be used to transform the synchronization of bit parity from a maximization problem into two 50/50 bit parity problems. We successfully tested the calibration procedure, but we currently can't test the proposed synchronization procedure as we have only one $45^o$ measuring device. Despite this we have successfully demonstrated control of the $45^o$ bit parity, so our next goal is to create a programmable quantum channel by merging a configurable quantum channel with a programmable classical channel. We have already simulated the interactions between a configurable quantum channel and a programmable classical channel, so our plan is to replace the simulated quantum channel with a real quantum channel.

\bibliography{references} 
\bibliographystyle{ieeetr}
%\bibliographystyle{unsrt}
%\bibliographystyle{plainnat}
%\bibliographystyle{plainurl}
%\printbibliography

\end{document}